\documentclass[english,a4paper,aps,prb,twocolumn,amsmath,amssymb,showpacs]{revtex4}
\usepackage{epsfig}
\usepackage{graphicx}
\usepackage{babel}
\usepackage{amsmath,amssymb}
\usepackage{color}
\usepackage[utf8]{inputenc}

\newcommand{\br}{\mathbf{r}}

\newcommand{\bq}{\mathbf{q}}
\newcommand{\bp}{\mathbf{p}}
\newcommand{\bk}{\mathbf{k}}

\newcommand{\ms}{\text{S}}
\newcommand{\vs}{\mathbf{S}}

\newcommand{\tD}{\widetilde{\Delta}}

\newcommand{\Neel}{\text{N}{\acute{e}}\text{el}}
\newcommand{\zncroA}{\text{Zn}_{1-x}\text{Cd}_x\text{Cr}_2\text{O}_4}
\newcommand{\ymoo}{\text{Y}_2\text{Mo}_2\text{O}_7}

\begin{document}

\title{Spin glass transition in geometrically frustrated antiferromagnets with weak disorder}

\author{A. Andreanov, J.T. Chalker, T. E. Saunders, and D. Sherrington}
\affiliation{Theoretical Physics, University of Oxford, 1 Keble Road, Oxford, OX1 3NP, United Kingdom}

\begin{abstract}
We study the effect in geometrically frustrated antiferromagnets of weak, random variations in the strength of exchange interactions. Without disorder the simplest classical models for these systems have macroscopically degenerate ground states, and this degeneracy may prevent ordering at any temperature. Weak exchange randomness favours a small subset of these ground states and induces a spin-glass transition at an ordering temperature determined by the amplitude of modulations in interaction strength.  We use the replica approach to formulate a theory for this transition, showing that it falls into the same universality class as conventional spin-glass transitions. In addition, we show that a model with a low concentration of defect bonds can be mapped onto a system of randomly located pseudospins that have dipolar effective interactions. We also present detailed results from Monte Carlo simulations of the classical Heisenberg antiferromagnet  on the pyrochlore lattice with weak randomness in nearest neighbour exchange.
\end{abstract}

\date{\today}


\maketitle

\section{Introduction}

Frustration refers to competition between few-body interactions which hinders simple macroscopic long-range ordering.  In many systems it involves competition between ferromagnetic and antiferromagnetic interactions. However, frustration can also occur in systems with purely antiferromagnetic interactions. Geometrically frustrated antiferromagnets\cite{GFAFMreviews} constitute a large class of materials in which the frustration has a purely structural origin and gives rise to highly degenerate ground states. In some instances a consequence of this degeneracy is that the system has no ordered, low-temperature phase, instead remaining in the paramagnetic phase down to zero temperature. 

It is well known that disorder in the form of quenched random-signed few-body interactions leads in high enough dimensions to a spin glass phase separated from the high temperature paramagnet by a true, if unusual, phase transition.\cite{binder1986} It has been a longstanding question whether the addition of  random interactions on top of those of a geometrically frustrated antiferromagnet could also lead to a true spin glass phase at low temperatures,\cite{villain} and if so, whether and to what extent the transition and low temperature phase are similar to those of conventional spin glasses.

In fact, spin glass like freezing has been observed in a number of geometrically frustrated magnets. These materials characteristically have a Curie-Weiss constant $\theta_{CW}$ of magnitude much greater than the freezing temperature $T_{\rm f}$. Some examples are {$\rm{SrCr_{8}Ga_{{4}}O_{19}}$}  ($\theta_{CW}\simeq -500K, T_F\simeq 4K$),~\cite{Ramirez1990,Ramirez1992,Martinez1992,Lee1996} 
$\ymoo$ (${\theta}_{CW}\simeq -200K$, $T_\text{f}\simeq 22K$),~\cite{Gingras1997,Gardner1999,Booth2000} and $\zncroA$ for $x \simeq 0.05$ ($\theta_{CW}\simeq -390, T_{\rm f} \simeq 12K$).~\cite{Broholm2000,Broholm2002} Typical observed features include differences between field-cooled and zero-field-cooled susceptibilities below $T_{\rm f}$, and in some cases an increase in non-linear susceptibility $\chi_\text{nl}$ close to $T_{\rm f}$.  In particular,  the existence of a sharp spin glass transition has been rather clearly established through detailed experiments on $\ymoo$.~\cite{Gingras1997,Gardner1999}

The reason for the observed freezing has long been a puzzle. On one hand, it has been established that simple models without disorder do not show freezing.\cite{Chalker-Moessner1998} On the other hand, samples exhibiting spin glass order either contain little structural disorder~\cite{Gingras1997} that could be invoked to explain the transition, or have a transition temperature that does not correlate straightforwardly with the level of the identified form of disorder. Indeed, in {$ \rm{SrCr_{8-x}Ga_{{4+x}}O_{19}}$} the transition temperature decreases with increasing disorder, as represented by the composition $\rm x$.\cite{Ramirez1990,Martinez1992}

A possible origin for a low temperature spin glass phase in frustrated magnets is suggested by recent experiments that show the importance of random strains in the samples.  Such strains, via magneto-elastic coupling,\cite{Broholm2000} generate randomness in the strength of antiferromagnetic exchange and hence may account for a spin glass phase at low temperatures. In the material  $\ymoo$, disorder in $\text{Mo}-\text{Mo}$ distances has been detected using XAFS.\cite{Booth2000} Separately, in $\zncroA$ disorder can be introduced in a controlled fashion by varying the composition $x$. Since Zn$^{2+}$ and Cd$^{2+}$ have different ionic radii, this non-magnetic disorder is expected to introduce random strains. Moreover, the fact that the undoped material ($x=0)$ has a low-temperature phase transition at which a frustration-relieving lattice distortion and N\'eel order appear together \cite{Broholm2000} suggests there is significant magnetoelastic coupling. It is therefore striking that small disorder levels ($x \gtrsim 0.03$) give rise to spin glass order at low temperature in place of the N\'eel phase. Disorder in the strength of exchange interactions, induced by distortions generated around \text{Cd} sites,\cite{Broholm2002} seems a likely origin for this behaviour.

Against this background, our aim in this paper is to study spin glass ordering in geometrically frustrated antiferromagnets with weak exchange randomness. Earlier studies of model frustrated systems in which low levels of disorder induce a spin glass phase are reported in Refs.~\onlinecite{campbell} and \onlinecite{berker},  and an earlier investigation by others of the problem we consider is described in Ref.~\onlinecite{bellier-castella2001}. A short account of some of our work has been given  in a previous publication~\cite{Saunders2007} by two of the authors. Here we present extended results, including a mapping to an analogue of the conventional spin glass theory. We show, both analytically and from simulations, that weak exchange randomness indeed generates spin glass order, with a transition temperature proportional to the amplitude of exchange randomness, albeit with a different proportionality constant than in a conventional system without the geometrical frustration. As a result of the dominant, average antiferromagnetic exchange, thermal fluctuations near the spin glass transition temperature are highly constrained: they lie within the ground state manifold of the equivalent system without quenched disorder. 
The ground states of the disorder-free system are macroscopically degenerate and have power-law correlations. Restriction of fluctuations to this set of states has two consequences. First, randomness in nearest neighbour exchange generates effective interactions that are long ranged, though not sufficiently so to change the universality class for critical behaviour. Second, the transition temperature arising from a given amplitude of disorder is higher in a geometrically frustrated system than it would be if the mean exchange interaction were zero.

An outline of the paper is as follows: in Section \ref{th} we introduce and discuss the replica treatment of a frustrated antiferromagnet with weak, homogeneous exchange randomness; in Section \ref{dilute} we consider dilute disorder, for which we map the geometrically frustrated magnet to a set of  pseudospins with random dipolar interactions; in Section \ref{ns} we present results from numerical simulations of the Heisenberg antiferromagnet on the pyrochlore lattice with exchange randomness; in Section \ref{fss} we use finite size scaling to analyse these results; and in Section \ref{conclusions} we make concluding remarks.

\section{Replica Theory}
\label{th}

\subsection{Setting the stage}
\label{th-ss}

Our starting point is a pure system consisting of classical $m$-component spins on the pyrochlore lattice with antiferromagnetic interactions, described by the Hamiltonian $\mathcal{H}_0=\sum_{\langle ij\rangle} J\,\vs_i\cdot\vs_j$. Here the sum $\sum_{\langle ij\rangle}$ runs over pairs of nearest neighbours, and spins are of unit magnitude. 
This model is geometrically frustrated and has a macroscopically degenerate ground state. As a consequence, ordering is suppressed and the system remains paramagnetic down to zero temperature~\cite{Chalker-Moessner1998} (except for the case $m=2$, in which thermal fluctuations induce collinear order at low temperatures,\cite{Chalker-Moessner1998} and which we exclude in the following). Another feature of the model is the emergence of power-law spin correlations at temperatures $T\ll J$, \cite{Isakov2004} and one of our concerns is to understand how these correlations influence spin glass phenomena in frustrated magnets.

We introduce exchange disorder by replacing $J$ in $\mathcal{H}_0$ with $J_{ij}=J+\delta J_{ij}$, where $\delta J_{ij}$ is random and has zero mean.  We take the distribution of  $\delta J_{ij}$ to be Gaussian with variance $\tD^2$ in our analytic work, and uniform on $[-\Delta,\Delta]$ in our numerical simulations, setting $\tD=\Delta/\sqrt{3}$ so that the variance is the same in both cases.
The regime of interest is $\Delta \ll J$; the opposite situation, $\Delta \gg J$, describes a system where disorder dominates over the antiferromagnetic coupling and a conventional spin glass phase is expected at low temperatures.

We hence consider the Hamiltonian
\begin{equation}
\mathcal{H}=J\sum_{\langle ij\rangle}\vs_i\cdot\vs_j+\sum_{\langle ij\rangle}\delta J_{ij}\vs_i\cdot\vs_j\,.
\end{equation}
This model is of direct physical interest in both the Heisenberg and Ising cases, the latter arising as a description of spin ice materials in which strong single ion anisotropy constrains spin orientations.

\subsection{Spin ice}
\label{th-ice}

We first consider Ising spins, for which $\ms_i=\pm 1$; the vector case is treated in Sec.~\ref{th-vec}. We use the replica trick~\cite{Edwards-Anderson1975} to carry out the disorder average. This produces an effective replica Hamiltonian
\begin{equation}
\label{th-ice-hav}
\beta\mathcal{H}_\text{av}=\frac{J}{T}\sum_{\langle ij\rangle,a}\ms_i^a\ms_j^a-\frac{\tD^2}{T^2}\sum_{\langle ij\rangle,(ab)}\ms_i^a\ms_i^b\ms_j^a\ms_j^b,
\end{equation}
where the replica labels take values $a,b=1,...n$, the summation $\sum_{(ab)}$ is over $a<b$, and  the limit $n\to 0$ is to be taken at the end of the computation. The last term can be rewritten as
\begin{equation}
-\frac{\tD^2}{2T^2}\sum_{ij,(ab)}\ms_i^a\ms_i^bK_{ij}\ms_j^a\ms_j^b,
\end{equation}
where $K_{ij}$ is the adjacency matrix on the pyrochlore lattice. The factor $1/2$ accounts for double counting in summation over sites.

Next we introduce local overlaps $Q_{ab}^i$ defined on sites of the lattice via a Hubbard-Stratonovitch transformation~\cite{Hubbard-Stratonovitch} which decouples the four-spin term, yielding\cite{footnote}
\begin{equation}
\beta\mathcal{H}_\text{Q}=\sum_{(ab)}\left\{\frac{1}{2}\sum\limits_{ij}Q_{ab}^i K_{ij}^{-1}Q_{ab}^j-\frac{\tD}{T}\sum_i Q_{ab}^i\ms_i^a\ms_i^b\right\}\,.
\end{equation}

The overall partition function can now be written as an integral on $Q_{ab}^i$ of $\exp[-\beta\mathcal{H}_\text{eff}]$, where the effective Hamiltonian is given by
\begin{equation}
\beta\mathcal{H}_\text{eff}=\frac{1}{2}\sum_{(ab),ij}Q_{ab}^iK_{ij}^{-1}Q_{ab}^j-W_{\tD}[Q]\,,
\end{equation}
with
\begin{equation}
W_{\tD}[Q]=\log\left\langle e^{\frac{\tD}{T}\sum_{(ab),i}Q_{ab}^i\ms_i^a\ms_i^b}\right\rangle_0\,.
\end{equation}
Here $\langle\cdots\rangle_0$ denotes an average with respect to the clean system Hamiltonian, $\mathcal{H}_0$. To advance further we need to evaluate $W_{\tD}[Q]$. An exact calculation is not possible and so we rely on a high-temperature expansion of $W_{\tilde{\Delta}}$.~\cite{Bray-Moore1979} That is, we expand the exponential in powers of $\tD/T$. Since $\mathcal{H}_0$ does not couple different replicas, averages $\langle\cdots\rangle_0$ with distinct replica labels factorise into products of averages with the same replica labels. Averages over odd numbers of spins vanish due to the absence of ordering in the pure system. As a consequence there is no term linear in $Q$. At order $(\tD/T)^2$ and $(\tD/T)^3$ only two-spin correlators $G_{ij}=\langle\ms_i\ms_j\rangle_0$ appear in the expansion. At third order this is due to the fact that bare correlators with an odd number of spin vanish, and those with an even number of spins greater than two are excluded due to constraints on replica indices that prohibit having four or all replica labels equal. 
Truncating the expansion at third order, we obtain
\begin{widetext}
\begin{gather}
W_{\tD}[Q]=\frac{\tD^2}{2T^2}\sum_{(ab),ij}Q_{ab}^i G_{ij}^2Q_{ab}^j + \frac{\tD^3}{T^3}\sum_{(abc),ijk}V_{ijk}Q_{ab}^i Q_{ac}^j Q_{bc}^k\,,\qquad {\rm where} \qquad
V_{ijk}=G_{ij}G_{ik}G_{jk}\,.
\end{gather}
\end{widetext}

It is convenient to split $\beta H_\text{eff}$ into $S_2$, the part quadratic in $Q$, and $S_\text{int}$, an interaction part. Fourier transformation block diagonalises $S_2$. Since the pyrochlore lattice has fours sites per unit cell, Fourier components carry both a wavevector $\bq$ and a sublattice label $\kappa$. We use a hat to denote the Fourier transform of a function, and write the spin-spin correlator of the pure system as $\widehat{G}_{\kappa\eta}(\bq)=\langle\ms_\kappa(-\bq)\ms_\eta(\bq)\rangle_0$. We also introduce the shorthand $\widehat{G^2_{\kappa\eta}}(\bq)=\int_\bp\widehat{G}_{\kappa\eta}(\bq-\bp)\widehat{G}_{\kappa\eta}(\bp)$. With this notation we have

\begin{widetext}
\begin{eqnarray}
\label{th-ice-s2}
S_2[Q] & =  & \frac{1}{2}\sum_{(ab),\kappa\eta}\int_\bq\widehat{Q}_{ab}^\kappa(-\bq)\left[\widehat{K}^{-1}_{\kappa\eta}(\bq)-\frac{\tD^2}{T^2}
\widehat{G^2_{\kappa\eta}}(\bq)
\right]\widehat{Q}_{ab}^\eta(\bq)\\
\label{th-ice-int}
{\rm and} \qquad S_\text{int}[Q] & = & -\frac{\tD^3}{T^3}\sum\limits_{(abc),\kappa\eta\omega}\int_{\bq\bp}V_{\kappa\eta\omega}(\bq,\bp)\widehat{Q}_{ab}^\kappa(\bq)\widehat{Q}_{ac}^\eta(\bp)\widehat{Q}_{bc}^\omega(-\bp-\bq)\,,\\
{\rm where} \qquad V_{\kappa\eta\omega}(\bq,\bp) & = & \widehat{G}_{\kappa\eta}(\bq)\widehat{G}_{\kappa\omega}(\bp)\widehat{G}_{\eta\omega}(-\bq-\bp)\,.
\end{eqnarray}
\end{widetext}


A phase transition is signalled  by the vanishing at the mean field critical temperature $T_c^\text{MF}$ of one of the eigenvalues of the kernel of $S_2$. Physically, one expects the critical mode to have wavevector zero. 
Because of translational invariance, a spatially uniform vector $Q_{ab}^i=\varphi_{ab}\cdot v$ is an eigenvector of the kernel 
at all temperatures. This mode is also separately an eigenvector of $\mathbf{K}(0)$, with an eigenvalue that we denote by $\lambda_4(0)$,
and of $\widehat{G_{\kappa\eta}^2}(0)$.
The temperature $T_c^\text{MF}$ is therefore determined by the equation
\begin{equation}
\label{th-ice-tc}
\frac{1}{\lambda_4(0)}-\frac{1}{4}\left(\frac{\tD}{T_c^\text{MF}}\right)^2\sum\limits_{\kappa\eta}
\widehat{G^2_{\kappa\eta}}(0)
=0\,.
\end{equation}

In the regime of most interest, $\Delta\ll J$, the temperature scale $J$ below which dipolar spin correlations  develop is much larger than the spin glass transition temperature. For temperatures $T \sim T_c^\text{MF}$ spin fluctuations are effectively confined to the groundstate manifold of the disorder-free system, and disorder acts as a perturbation. Under these conditions we can use the results of Refs.~\onlinecite{Isakov2004} and \onlinecite{Garanin-Canals1999} to write the two-spin correlation function approximately in terms of the normalised eigenvectors $\{U_{\xi=1,2,3,4}\}$ of the adjacency matrix $\widehat{\mathbf{K}}$. This matrix has two flat bands, with eigenvalues $\lambda_{\xi=1,2}=-2$ that are independent of wavevector, and two dispersive bands, with eigenvalues  $\lambda_3(\bq)<\lambda_4(\bq)$. We have\cite{Isakov2004,Garanin-Canals1999}
\begin{gather}
\label{th-ice-g}
\widehat{G}_{\kappa\eta}(\bq)=2\sum_{\xi=1,2}(U_\xi)_\kappa(\bq)(U_\xi)_\eta(\bq)\,.
\end{gather}

This approximation and the orthonomality of the eigenvectors $\{U_\xi\}$ yields $\sum_{\kappa\eta}\widehat{G_{\kappa\eta}^2}(0)=8$. Now $\lambda_4(0)=6$ (see Ref.~\onlinecite{Isakov2004}), and  so we obtain from Eq.~\eqref{th-ice-tc} the result $T_c^\text{MF}=\tD\sqrt{12}$. A conventional mean-field result for $T_c^\text{MF}$ is also contained in Eq. \eqref{th-ice-tc} when one sets $J=0$. In this case $\widehat{G}_{\kappa\eta}(\bq)=\delta_{\kappa\eta}$, which implies $\sum_{\kappa\eta}\widehat{G_{\kappa\eta}^2}(0)=4$ and $T_{c,J=0}^\text{MF}=\tD\sqrt{6}$. Thus, within the framework of our calculation, there is an increase in the value of $T_c^\text{MF}$ by a factor of $\sqrt{2}$ due to the correlations arising from uniform 
frustration.\cite{footnote2}
The increase in the spin glass phase transition temperature compared with that of a conventional system with $J=0$ has a simple physical explanation: the geometrical frustration severely reduces the phase space available for spin-glass-destabilising thermal fluctuations in the low temperature manifold.

To describe the transition we retain only the branch of soft modes.  We denote the corresponding eigenvector of the kernel of $S_2$ by $v^\kappa(\bq)$, and the associated eigenvalue by $E(\bq)$.  Writing $Q_{ab}^\kappa(\bq)\approx\varphi_{ab}(\bq)v^\kappa(\bq)$, the effective theory for the spin glass transition is
\begin{gather}
S[\varphi]=\frac{1}{2}\sum\limits_{(ab)}\int_\bq E(\bq)|\varphi_{ab}\bq)|^2-\\
\left(\frac{\tD}{T_c^\text{MF}}\right)^3\sum\limits_{(abc)}\int_{\bq\bp}V(\bq,\bp)\varphi_{ab}(\bq)\varphi_{ac}(\bp)\varphi_{bc}(-\bq-\bp)\notag\,,
\label{th-ice-v-keo}
\end{gather}
where
\begin{equation}
V(\bq,\bp)=\sum\limits_{\kappa\eta\omega}V_{\kappa\eta\omega}(\bq,\bp)v^\kappa(\bq)v^\eta(\bp)v^\omega(-\bp - \bq)
\label{cubic}
\end{equation}
In conventional spin-glasses $E(\bq)\sim E(0)+A\bq^2$ as $\bq \to 0$, with  $A$ a positive constant.  If this behaviour persists in our case, the conventional effective theory is retrieved. Direct diagonalisation of $S_2$ for $\bq\neq0$ is complicated and we rely instead on non-degenerate perturbation theory to study the small $\bq$ behaviour of $E(\bq)$. To first order in $(\tD/T)^2$ the eigenvalue $E(\bq)$ is given by 
$$\frac{1}{\lambda_4(\bq)}-\left(\frac{\tD}{T}\right)^2\sum\limits_{\kappa\eta}U_4^\kappa(\bq)\int_\bp\widehat{G}_{\kappa\eta}(\bq-\bp)\widehat{G}_{\kappa\eta}(\bq)U_4^\eta(\bq).$$ 
We consider in turn the contributions to $E(\bq)-E(0)$ from each term in this expression.
First, using an explicit form for $\lambda_4(\bq)$~\cite{Isakov2004} it is easy to check that $\lambda_4^{-1}(\bq) - \lambda_4^{-1}(0)\sim A_0\bq^2$ as $\bq\to0$ with $A_0>0$. 
Next, direct evaluation of $\sum_\kappa [U_4^\kappa(\bq)- U_4^\kappa(0)]$ shows that it vanishes faster than $q^2$ as $\bq \to 0$. Analysis of the contribution from the second term therefore reduces to the evaluation of
\begin{equation}
\label{th-ice-qq}
-\left(\frac{\tD}{T}\right)^2 q_\alpha q_\beta\sum\limits_{\kappa\eta}\int_\bp\left[\nabla_\alpha\nabla_\beta\widehat{G}_{\kappa\eta}(\bp)\right]\widehat{G}_{\kappa\eta}(\bp)\,,
\end{equation}
where $\nabla_\alpha=\partial/\partial p_\alpha$. The integral on $\bp$ is most easily understood
in real space, where it takes the form
\begin{equation}
\label{th-ice-q2}
-\sum_{\br_i\br_j}(\br_i-\br_j)_\alpha(\br_i-\br_j)_\beta G^2(\br_i-\br_j).
\end{equation}
Crucially, the power law decay $G({\bf r}) \sim r^{-3}$ at large $r$ is fast enough that the sum in Eq.~(\ref{th-ice-q2}) converges, giving a finite result for $\alpha=\beta$ and zero by symmetry for $\alpha \not= \beta$. Moreover, 
the signs of Eqs.~\eqref{th-ice-qq} and \eqref{th-ice-q2} combine to give an overall positive sign to the coefficient of $\bq^2$. Taken together these results ensure that $E(\bq)\sim E(0)+A\bq^2$ with positive $A$.

Within the context of the effective theory it is also justified to replace $V(\bq,\bp)$ by its limit for $\bq,\bp\to0$. Taking this limit in Eq.~(\ref{cubic}) and using the asymptotic form for $\widehat{G}$ 
we find 
$$V(\bq,\bp)=(\widehat\bp,\widehat{\bk})^2+(\widehat\bq,\widehat{\bk})^2+(\widehat\bp,\widehat\bq)^2-(\widehat\bp,\widehat\bq)(\widehat\bp,\widehat{\bk})(\widehat\bq,\widehat{\bk})$$ 
where $\widehat{\bp}$,  $\widehat{\bq}$, and $\widehat{\bk}$ denote unit vectors in the directions of $\bp$, $\bq$ and $\bk \equiv \bp + \bq$, respectively.

Summarising these results, we have obtained an essentially conventional replica theory at mean field level. A critical theory, almost identical to the one derived for Edwards-Anderson model~\cite{Bray-Moore1979}, also follows. The only difference to the standard version is in the form of the interaction term, which has a non-trivial wave-vector dependence in our case, originating from dipolar correlations present in pure system. The effective critical theory is
\begin{gather}
\label{th-ice-th}
S[\varphi]=\sum_{(ab)}\int_\bq(\bq^2+\tau)|\varphi_{ab}|^2(\bq)+\\
+\frac{g}{6}\int_{\bq\bp}V(\bq,\bp)\sum_{(abc)}\varphi_{ab}(\bq)\varphi_{ac}(\bp)\varphi_{bc}(-\bp - \bq)\,,\notag
\end{gather}
where $\tau\sim 1-(T^\text{MF}_c/T)^2$ and  $g\sim(\tD/T_c^\text{MF})^3$.

\subsection{Vector spins}
\label{th-vec}

In this section we sketch the generalisation of the results derived above for the generic case of $m$-component spins.
The derivation follows closely that for the Ising case with minor modifications due to additional spin component labels. 
Local overlaps $Q_{ab}^{\alpha\beta}(i)$ now carry an additional pair of spin component labels $\alpha,\beta$ and summation over $ab$ is unrestricted. Also, at variance with the Ising case diagonal terms ($a=b$) now give non-constant contribution to the effective Hamiltonian and cannot be dropped. As a consequence one has to define diagonal overlaps $Q_{aa}^{\alpha\beta}(i)$, which contribute to the expansion of $W_{\tD}[Q]$. This in turn leads to appearance of four and higher spin correlators already at order three in the expansion of $W_{\tD}$. However it affects only the form of the cubic term, and not that of the quadratic term which determines the value of critical temperature. 
The mean-field critical temperature expression generalises to $T^\text{MF}_c=\tD\sqrt{12}/m.$ The critical theory is very similar to the spin ice case, being
\begin{gather}
\label{th-vec-th}
S[\varphi]=-\sum_{a\alpha}\varphi_{aa}^{\alpha\alpha}(0)+\frac{1}{2}\sum_{ab,\alpha\beta}\int_\bq(\bq^2+\tau)|\varphi_{ab}^{\alpha\beta}|^2(\bq)+\\
+\frac{g}{6}\int_{\bq\bp}V(\bq,\bp)\sum_{abc,\alpha\beta\gamma}\varphi_{ab}^{\alpha\beta}(\bq)\varphi_{ac}^{\alpha\gamma}(\bp)\varphi_{bc}^{\beta\gamma}(-\bp - \bq)\,.\notag
\end{gather}

To conclude, critical replica theories for disordered frustrated magnets at all values of $m$ (except, as indicated above, $m=2$) coincide with the conventional critical theory for spin-glasses.~\cite{Bray-Moore1979} The underlying geometrical frustration reveals itself in a remaining wave-vector dependence of the interaction vertex and an increased mean-field value of the critical temperature as compared to the conventional case. 

\section{Dilute Impurities}
\label{dilute}

It is also interesting to consider a model of dilute disorder, in which a low density of isolated tetrahedra have exchange interactions that are different for  different pairs of spins within the tetrahedron. We show in this section that each such tetrahedron has a pseudospin degree of freedom. We find that entropic interactions between these pseudospins, mediated by spins in the remaining tetrahedra, have a dipolar form. In these way we arrive at a similar conclusion to the one reached in Section \ref{th}, but for a different version of the problem and by a different route.

\begin{figure}
\centering
\epsfig{file=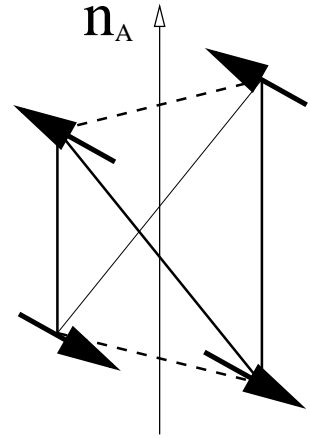,width=40mm}
\caption{\label{fig-tet} A ground state configuration of spins at vertices of an isolated tetrahedron in which there are antiferromagnetic exchange interactions of strength $J$ between pairs joined by solids lines, and of strength $J-\delta J$ (with $0< \delta J \ll J$) between pairs joined by dashed lines. The vector ${\bf n}_A$ is perpendicular to both of the links between ferromagnetically aligned pairs of spins }
\end{figure}

As a first step, consider a single tetrahedron taken from this lattice, with spins $\mathbf{S}_1 \ldots \mathbf{S}_4$ at the vertices. With all exchange interactions equal, its ground states are the configurations for which $\sum_i \mathbf{S}_i = \mathbf{0}$. The spin stiffness is zero in this toy problem in the sense that, within the set of ground states, the orientations of a pair of spins can be chosen arbitrarily. The consequences of fluctuations in $\delta J_{ij}$ with amplitude $\Delta$ have been set out in Ref.~\onlinecite{bellier-castella2001} and \onlinecite{tchernyshyov2002}: generically, a unique ground state is selected (up to global spin rotations) in which all four spins are collinear 
and the total spin of the tetrahedron is zero. In such a configuration
the four spins can be grouped into two ferromagnetically aligned pairs, and energy
is minimised by picking these pairs appropriately, as illustrated in Fig.~\ref{fig-tet}. Non-zero $\delta J_{ij}$ hence induce a ground state stiffness, since changes in the relative orientation of a pair of spins cost an energy ${\cal O}(\Delta)$. 

Extending this discussion, consider a pyrochlore lattice in which a randomly selected, dilute subset of special tetrahedra have interactions of unequal strength, while in the remainder all $\delta J_{ij}$ are zero. Provided dilution is sufficiently high, ground states are macroscopically degenerate, and in ground states each special tetrahedron has collinear spins at its vertices. The orientations of these quartets of collinear spins at different special tetrahedra are independent, and constitute some of the ground state degrees of freedom. We label the special tetrahedra by $A$ and specify these orientations with unit vectors $\boldsymbol{\sigma}_A$ in spin space. In addition, to characterise the realisation of quenched disorder we introduce unit vectors ${\bf n}_A$ in real space, defined to be perpendicular to both of the links on the lattice that join ferromagnetically aligned pairs of spins in tetrahedron $A$, as shown in Fig.~\ref{fig-tet}. (Since  ${\bf n}_A$  serves to define an axis,  $\pm{\bf n}_A$ are equivalent.) These vectors are each aligned along one of the cubic axes of the pyrochlore lattice and are quenched random variables. At temperatures $T\ll \Delta$ only the ${\bf n}_A$, and not the magnitudes of the $\delta J_{ij}$, are important to characterise the disorder. 

Integration over all other degrees of freedom induces an effective, entropic interaction between the $\boldsymbol{\sigma}_A$'s. The form of this effective interaction can be deduced by using the parameterisation of the ground states of the disorder-free model in terms of a gauge field, introduced in Refs.~\onlinecite{Isakov2004}. 

The essence of this parameterisation can be summarised as follows, treating in the first instance the case of Ising spins ${S}_i$. A three-component vector field $\mathbf{B}(\mathbf{r})$ is chosen to represent spin configurations, in such a way that the condition for a configuration to be a ground state is $\mathbf{\nabla}\cdot \mathbf{B}(\mathbf{r})=0$. At the microscopic level, this is achieved in two steps.\cite{Isakov2004} First, one notes that the centres of tetrahedra on the pyrochlore lattice themselves form a diamond lattice, which is bipartite. It is therefore possible to define a unit vector $\hat{e}({\bf r}_i)$ in real space at each site ${\bf r}_i$ of the pyrochlore lattice, with the orientation convention that it is directed from one chosen diamond sublattice towards the other.  Second, one defines $\mathbf{B}(\mathbf{r}_i) \equiv \hat{e}({\bf r}_i){S}_i$, which has zero lattice divergence in ground states. After coarse-graining, $\mathbf{B}(\mathbf{r})$ is treated as continuous, divergence-free field.
Extending these ideas to treat $m$-component spins, one introduces $m$ fields $\mathbf{B}^\alpha(\mathbf{r})$, with $\alpha=1 \ldots m$, related at the lattice level to spin components $S^\alpha_i$ by $\mathbf{B}^\alpha(\mathbf{r}_i) \equiv \hat{e}({\bf r}_i){S}^\alpha_i$.
The coarse-graining procedure gives rise to an entropic weight that favours configurations with small field strengths. Writing this weight as $e^{-S_0}$, $S_0$ is postulated\cite{Isakov2004} to have the form
\begin{equation}
S_0 = \frac{\kappa}{2}\int \text{d}^3\mathbf{r} \sum_\alpha |\mathbf{B}^\alpha(\mathbf{r})|^2
\end{equation}
where $\kappa$ characterises an entropic stiffness, which is distinct from the energetic stiffness that arises  when $\delta J_{ij}$ is non-zero for all nearest neighbour pairs.

In this language, the condition that the spins of a special tetrahedron located at a random position $\mathbf{r}_A$ are collinear with orientation $\boldsymbol{\sigma}_A$ translates into the condition on the fields $\mathbf{B}^\alpha(\mathbf{r})$ 
that $\mathbf{B}^\alpha(\mathbf{r}_A) = \sigma^\alpha_A \mathbf{n}_A$.
We impose these constraints by introducing three-component fields $\boldsymbol{\phi}^\alpha_A$ and using on the special tetrahedra
\begin{equation}
\delta\left(\mathbf{B}^\alpha(\mathbf{r}_A) - \sigma^\alpha_A\mathbf{n}_A\right)=\frac{1}{(2\pi)^3} \int \text{d}^3 \boldsymbol{\phi}^\alpha_A e^{i\boldsymbol{\phi}^\alpha_A\cdot (\mathbf{B}^\alpha(\mathbf{r}) - \sigma^\alpha_A\mathbf{n}_A)}\,.
\nonumber
\end{equation}
Integrating out the fields $\mathbf{B}^\alpha(\mathbf{r})$ and the variables $\boldsymbol{\sigma}_A$, we arrive at a weight $e^{-S_{\rm eff}}$ for $\boldsymbol{\phi}^\alpha_A$ of the form
\begin{equation}
S_{\rm eff} = \frac{1}{2\kappa}\sum_{\alpha\, AB}\boldsymbol{\phi}^\alpha_A\cdot\mathbf{M}(\mathbf{r}_A-\mathbf{r}_B)\cdot \boldsymbol{\phi}^\alpha_B+ \sum_{A} V(\varphi_A^2)\,.
\label{20}
\end{equation}
Here the $3 \times 3$ interaction matrix $\mathbf{M}(\mathbf{r})$ is dipolar, with elements $[\mathbf{M}(\mathbf{r})]_{kl} = (r^2\delta_{kl}-3r_kr_l)/4\pi r^5$. The one-body term $V(\varphi_A)$ is a function of $\varphi_A^2 = \sum_\alpha(\boldsymbol{\phi}^\alpha_A \cdot \mathbf{n}_A)^2$ and has the expansion $V(\varphi^2) = c_2\varphi^2+ c_4\varphi^4 \cdots$, where $c_2 = (2m)^{-1}$ and $c_4 =(4m^2[m+2])^{-1}$. This effective model has site disorder, since the tetrahedra labelled by $A$ and $B$ are selected at random. By this means, we have arrived at a model of randomly located interacting dipoles as a description in the $T\to 0$ limit. We expect the model to have a classical, zero-temperature phase transition between a paramagnetic phase at low density and a spin glass phase at high density. 

At finite temperature thermal excitations generate a finite correlation length $\xi$, which sets a maximum range for the interaction $\mathbf{M}(\mathbf{r})$. This correlation length diverges in the low temperature limit, exponentially in $J/T$ for the Ising spins, and as the power law $\xi \sim (J/T)^{1/2}$ for Heisenberg spins. On increasing temperature from zero in the spin glass phase, a transition to a paramagnet is expected, with a transition temperature set by the lower of two scales: one of these is the temperature at which $\xi$ becomes comparable to the spacing between defect tetrahedra; the other is the disorder strength $\Delta$. Despite the dipolar form of interactions at distances shorter than $\xi$ this transition is expected to be in the same universality class as with short-range exchange.\cite{Bray-Moore1982} 

In principle an analogous finite-temperature mapping could be made in terms of the model of Section II, 
by replacing the hard constraints ${\bf{B}}^{\alpha}({\bf{r}}_A)={\sigma}^{\alpha}_{A}{\bf{n}_A}$
with soft weightings $\exp(-E_{A}(\{J_{ij}^{A}\})/T)$. These will be determined by the energetic costs $E_{A}(\{J_{ij}^{A}\})$ of local spin-configurations drawn 
from the ground state manifold of the clean system, 
where $\{J_{ij}^{A}\}$ denotes the set of $\{J_{ij}\}$ within tetrahedron A. This mapping would yield an analogue of Eq.~(\ref{20}) with a summation over all A but with extra quench-random local $\phi$ weightings, again leading to spin glass behaviour as expected for a site-disordered Ginzburg-Landau spin glass.

\section{Numerical Simulations}
\label{ns}

In order to investigate our ideas further we turn to Monte Carlo simulations of the classical Heisenberg antiferromagnet with nearest neigbour exchange on the pyrochlore lattice. Our focus is on the effects of weak randomness in the strength of exchange interactions.  This section provides a complete description of work presented briefly in an earlier publication.~\cite{Saunders2007}

\subsection{Model and Method}

We take the exchange interaction between spins at neighbouring sites $i$ and $j$ to have strength $J+\delta J_{ij}$ with $\delta J_{ij}$ an independent random variable for each bond, uniformly distributed in the range $[-\Delta, \Delta]$. Our interest is in the limit $\Delta \ll J$ and our most extensive results are for $\Delta = 0.1J$. 

We simulate a sample in the shape of a rhomboid that has edges parallel to the primitive basis vectors of the lattice, with periodic boundary conditions between opposite faces. System size is specified by the linear dimension $L$ of the sample. The number of primitive unit cells in such a sample is $N=L^3$ and the total of number of spins is $N_{\rm s}=4L^3$. We present data for sample sizes in the range from $L=2$ to $L=7$. At each sample size it is necessary to average over different disorder realisations. We used $10^3$ realisations for $L=2$ but found that $200$ realisations are sufficient for $L=7$.  

We employ parallel tempering \cite{parallel} to ensure equilibration of large systems at low temperature. In this approach,  one simulates $N_{\rm array}$ copies of the system simulaneously. Each copy is at a different temperature, taken in a range from $T_{\rm min}$ to $T_{\rm max}$ with geometric spacing. This range is required to be wide, since $T_{\rm min}$ must be below the spin glass transition temperature while $T _{\rm max}$ must be high enough that relaxation at that temperature is fast. At the same time, adjacent temperatures should be sufficiently close that there is a high probability for configurations to be exchanged between them under the moves of the parallel tempering algorithm. This requires a sufficiently large value value of $N_{\rm array}$. We take  \cite{parallel}
\begin{equation}
N_{\rm array} = N_{\rm s}^{1/2} \ln(T_{\rm max}/T_{\rm min})\,.
\end{equation}

\subsection{Testing Equilibration}

\begin{figure}[tb]
\begin{center}
\includegraphics[height=50mm,width=80mm]{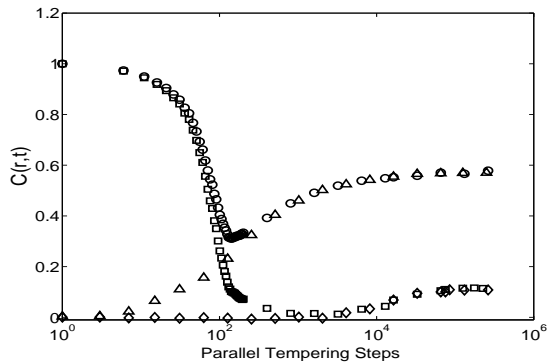}
\caption{\label{fig:equil} Evolution of the spin glass correlation function for two different starting configurations. Simulations starting from a random initial condition (for $r=1$ ($\triangle$), $r=r_{\rm max}$ ($\diamond$)) are compared with simulations starting in a $\Neel$ ordered state (for spatial separations $r=1$ ($\circ$), $r=r_{\rm max}$ ($\square$)). Error bars are omitted for clarity: in the worst case ($r=r_{\rm max}$) they are $\sim 10\%$.}
\end{center}
\end{figure}

We estimate equilibration times by studying the evolution of observables starting from different initial states. Two simple choices of initial state are an infinite temperature configuration with random spin orientations, and a N\'eel ordered configuration with collinear spins, which is a ground state of the model without exchange randomness. For each of these initial states we show in Fig.~\ref{fig:equil} the evolution with Monte Carlo time of the spin glass correlation function $C({\bf r})$ [defined in Eq.~(\ref{sgcf})] for $r=1$ and $r=r_{max}$, the maximum separation in a sample of size $L=7$, taking $\Delta/J=0.1$ and $T/\Delta=0.1$. This is the largest lattice size and lowest temperature, and hence the most difficult case, for which we present detailed results in our study of the spin glass transition. As seen from Fig.~\ref{fig:equil}, the equilibration time in this case is $\sim 10^5$ parallel tempering steps, although memory of the difference between initial configurations is lost after a shorter time.  When using parallel tempering it is also important that each copy of the system should visit every temperature simulated with equal probability.  We have checked that the simulation time for a copy to loose memory of its initial temperature is shorter than the equilibration time for the spin glass correlation function. On the basis of these tests, for $L=7$ we collect data after an equilibration time of $10^5$ parallel tempering steps. For smaller system sizes equilibration is more rapid.

Our equilibration times are similar to those for conventional Heisenberg spin glasses on the cubic lattice\cite{conventional}. It is worth noting that these equilibration times are smaller than those typically needed for Ising spin glasses.  Lee and Young suggest that this is because energy barriers are smaller in the Heisenberg model.\cite{conventional}  The extra degrees of freedom in the Heisenberg model mean that the simulation can find paths around energy barriers, rather than over them as in Ising systems.
 
\subsection{Specific Heat}

The low temperature heat capacity $C_{v}$ of the classical Heisenberg antiferromagnet  on the pyrochlore lattice is interesting as a diagnostic for macroscopic ground state degeneracy.\cite{Chalker-Moessner1998} Without degeneracy, equipartition and the fact that each spin has two degrees of freedom would give a classical low temperature heat capacity of $k_{\rm B}$ per spin. The smaller measured value of $(3/4) k_{\rm B}$ per spin demonstrates that one quarter of the degrees of freedom in the model make no contribution to $C_{v}$ because they can fluctuate without energy cost. We expect exchange randomness to eliminate this macroscopic number of zero modes, leaving only  the three zero modes associated with global spin rotations. The limiting low temperature value of the heat capacity per spin should then be $(1-3/2N_s)k_{\rm B}$. At higher temperatures the heat capacity is expected to have a broad maximum in the vicinity of the spin glass transition, and to remain finite and smooth even in the thermodynamic limit. 

\begin{figure}[tb]
\begin{center}
\includegraphics[height=50mm,width=80mm]{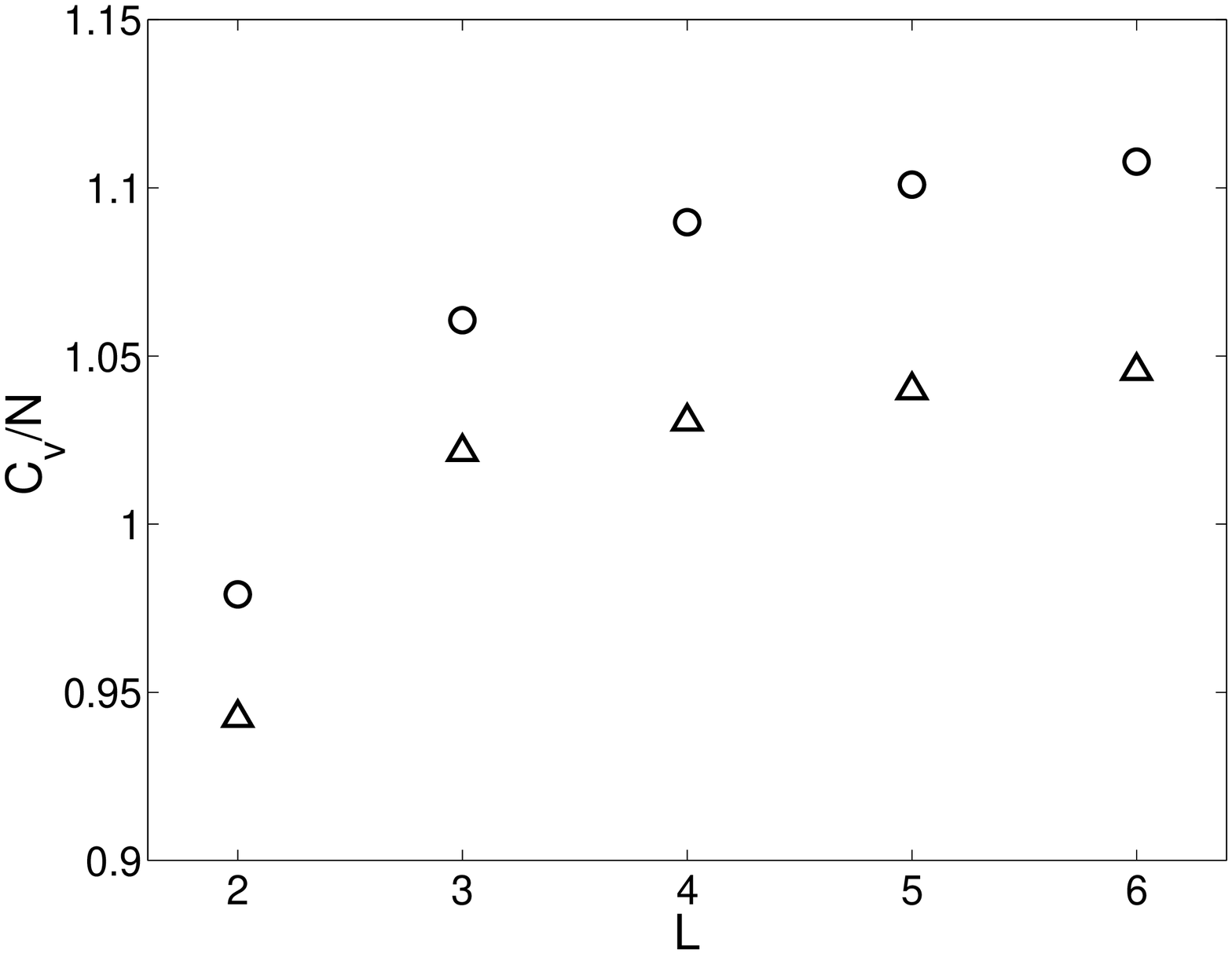}
\caption{\label{figure2} $C_v$ against $L$ for two temperatures: that of the maximum ($\circ$); and  $T/{\Delta}=0.1$ ($\triangle$), the lowest temperature simulated for larger system sizes.} 
\end{center}
\end{figure}

In our simulations  we determine the heat capacity from the variance of energy fluctuations. As reported previously, \cite{Saunders2007} the heat capacity has a broad maximum around $T/\Delta\sim0.45$ for $\Delta/J=0.1$, while at low temperatures it tends to unity for large system sizes. We show in Fig.~\ref{figure2} the dependence on system size of $C_v$ at the temperature of the maximum and at the low temperature $T/\Delta = 0.1$. As expected, both values approach a constant with increasing system size, which in the second case is close to $k_{\rm B}$. Further calculations down to $T/\Delta=10^{-2}$ for $L=4$ (not shown) confirm the expected value $(1-3/2N_s)k_{\rm B}$ more precisely.

\subsection{Spin Glass Correlation Function}

To search for spin freezing we study the spin glass correlation function $C({\bf r})$. This is defined in terms of behaviour in two copies of the system with identical disorder. Denoting thermal averages in copies 1 and 2 by $\langle \ldots \rangle_1$ and  $\langle \ldots \rangle_2$, and the disorder average by $[ \ldots ]_{av}$, we have
\begin{equation}
\label{sgcf}
C({r}) = \left[\langle \mathbf{S}(0)\cdot\mathbf{S}(\mathbf{r})\rangle_1 \langle \mathbf{S}(0)\cdot\mathbf{S}(\mathbf{r})\rangle_2\right]_{av}\,.
\end{equation}

\begin{figure}[tb]
\begin{center}
\includegraphics[height=50mm,width=80mm]{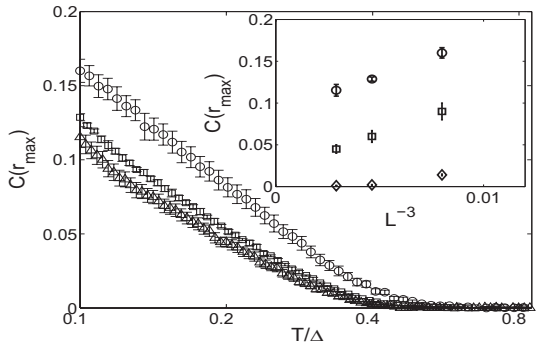}
\caption{\label{figure3} $C(r_{max})$ versus temperature for $L=7 (\triangle), L=6 (\square), \textrm{ and } L=5 (\circ)$. The inset shows the scaling of $C(r_{max})$ with system system size for $T/{\Delta}=0.1 (\circ)$, $T/{\Delta}=0.2 (\square)$, and $T/{\Delta}=0.4 (\diamond)$.}

\end{center}
\end{figure}

Spin freezing is indicated by a non-zero limiting value for $C(r)$ at large $r$. We show in Fig.~\ref{figure3} the temperature dependence of  $C(r)$ for the maximum spin separation ($r=r_{\rm max}$)  in the three largest system sizes studied ($L=5,6,7$).  There is a clear transition within the temperature range $0.2<T/{\Delta}<0.4$. This behaviour is in marked contrast to that of the pure system, where $C(r)$ falls with $r$ as $r^{-6}$ in the low temperature limit, and exponentially at finite temperature. \cite{Isakov2004} 
   
The inset to Fig.~\ref{figure3} suggests that $C(r_{\rm max})$ tends to a finite constant below $T_f$ for large $L$, further supporting our conclusion that there is a finite temperature transition in the infinite system when weak bond disorder is present. The behaviour of the correlation function puts simple bounds on the transition temperature, $0.2\lesssim T_f/{\Delta}  \lesssim 0.4$. 

The behaviour of the spin glass correlation function thus provides convincing evidence for a spin glass transition in the model.  However, the value of the critical temperature has a large uncertainty.  As we discuss in the next section, finite size scaling can be used to sharpen the estimate for the critical temperature and to find approximate values for critical exponents.

\section{Finite-Size Scaling}
\label{fss}

There is extensive past work on finite size scaling analysis of the transition for conventional models of spin glasses that have zero mean exchange interaction.\cite{conventional} From this it has emerged that study of the behaviour of the spin glass correlation length is a particularly effective approach. We find that the situation is different in the case of geometrically frustrated systems. These have strong short-range correlations at low temperature, even in the absence of disorder, which complicate a scaling analysis using the limited range of system sizes available. Instead we employ scaling collapse of the spin glass susceptibility [Eq.~(\ref{chi-sg})] to obtain simultaneous estimates of $T_{\rm f}$, $\nu$ and $\gamma$.  We emphasise that our aim here is not to find precise values for the critical exponents; rather, our principal objective is to confirm that there is a finite temperature second order phase transition and determine its temperature.

The spin glass susceptibility, which is related to the non-linear susceptibility  $\chi_{nl}$, is defined as \cite{binder1986}
\begin{equation}\label{chi-sg}
\chi_{SG} = \sum_{\mathbf{r}}C(\mathbf{r}) \,.
\end{equation}
In the paramagnetic phase $\chi_{SG} \sim \mathcal{O}(1)$ since the only significant contribution to $C(r)$ is from small $r$. Its divergence at finite temperature signals a phase transition. Approaching the critical temperature from above, we expect
\begin{equation}
\label{chisg}
\chi_{SG} \sim t^{-\gamma} \,,
\end{equation}
where $t=(T-T_{\rm f})/T_f$ and $\gamma$ is the corresponding critical exponent. Eq.~(\ref{chisg}) holds if the dimension $d$ of the system is greater than $d_l$, the lower critical dimension, so that $T_{\rm f} > 0$. The evidence from the previous section strongly favours this scenario.  Furthermore, from the results of Section~\ref{th} we expect the upper critical dimension for the transition to have its conventional value, which is believed to be six\cite{binder1986}. The hyperscaling relations should therefore hold, and using $d\nu=\gamma+2\beta$ we can deduce the value of $\beta$ from scaling of $\chi_{SG}$.

\begin{figure}[tb]
\begin{center}
\includegraphics[height=50mm,width=80mm]{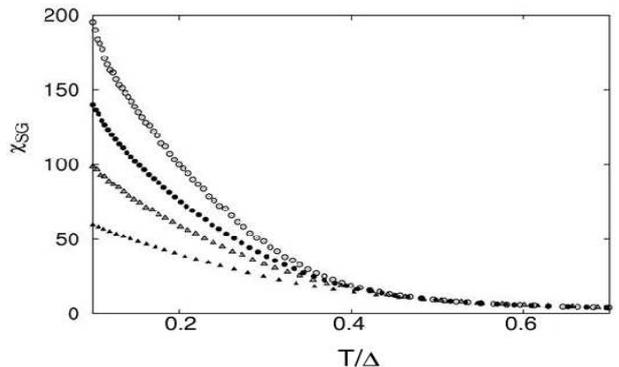}
\caption{\label{figure4} The spin glass susceptibility, $\chi_{SG}$ vs $T/\Delta$ for system sizes from $L=4$ (closed triangles) to $L=7$ (open circles).}
\end{center}
\end{figure}

If there is a thermodynamic phase transition then  we anticipate close to $T_f$ in a finite system the scaling behaviour
\begin{equation}
\label{chiscale}
\chi(T,L) = L^{\gamma/\nu}f(L^{1/\nu}t) \,,
\end{equation}
where $\nu$ is the critical exponent of the correlation length. The scaling function obeys $f(0)>0$ and $f(x)\sim x^{-\gamma}$ for $x \rightarrow \infty$.  In Fig.~\ref{figure4} we show $\chi_{SG}$ as a function of $T/{\Delta}$ for system sizes from $L=4$ to $L=7$. 

The rapid increase in $\chi_{SG}$ below $T/\Delta\sim 0.4$ provides clear evidence of a spin glass transition.  The scaling analysis of $\chi_\text{SG}$ is complicated by large finite-size effects. Due to the ground state constraint in frustrated antiferromagnets, our model has significant short-ranged correlations that are approximately independent of  system size. For $L=2$ and $L=3$ the contribution to $\chi_\text{SG}$ from these local correlations is significantly greater than the contribution from $C(r)$ at large distances and so these system sizes cannot be included in the analysis.

\begin{figure}
\centering
\epsfig{file=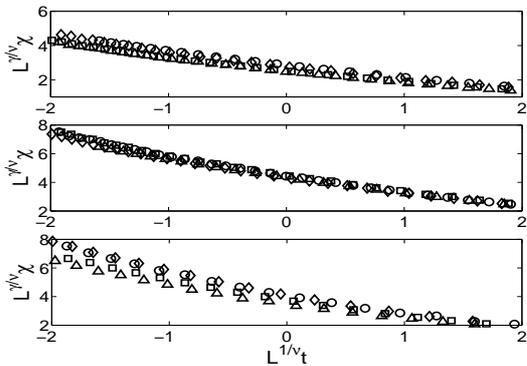,height=50mm,width=80mm}
\caption{\label{figure5} Scaling collapse of $\chi_{SG}$ for: (top panel)  $\nu=1.2, \gamma=1.2, T_c/{\Delta}=0.2$; (middle panel) $\nu=1.0, \gamma=1.45, T_c/{\Delta}=0.23$; (bottom panel) $\nu=0.9, \gamma=1.6, T_c/{\Delta}=0.32$}
\end{figure}

A scaling collapse of $\chi_{SG}$ for the system sizes $4\leq L \leq 7$ yields the best-fit parameters $T_{\rm f}/{\Delta} = 0.23(9)$, $\nu = 1.0(2)$ and $\gamma = 1.45(45)$ \cite{Saunders2007}. It is difficult to make detailed estimates of the errors in these values, but a simple approach is to explore the range of parameters that still gives reasonable data collapse. In Fig.~\ref{figure5} the scaling collapse for the best fit  is compared with behaviour for two  `worst-case' fits: (i)  $T_{\rm f}/{\Delta}=0.2$, $\nu=1.2$ and $\gamma=1.2$; and (ii) $T_{\rm f}/\Delta = 0.32$,  $\nu=0.9$ and $ \gamma=1.6$. These worst-case fits were generated by fixing the value of $T_{\rm f}$ and then adjusting the values of $\nu$ and $\gamma$ to minimise scatter of the data. As the data collapse is visibly poorer for both the worst cases, we believe they set bounds on the value of $T_{\rm f}$. In addition,  since the exponent values that produce good scaling collapse are correlated with the fitting value for $T_{\rm f}$, they give bounds on $\nu$ and $\gamma$. These are the uncertainties quoted above.

Our results can be compared with those from simulations on conventional spin glass models, and with experiment. As a first step, it is interesting to examine the value of $T_{\rm f}$. The most direct comparison would be between $T_{\rm f}/\Delta$ at large $J$ and at $J=0$, evaluated in both cases for the Heisenberg model on the pyrochlore lattice. Because we do not have data for this  lattice at  $J=0$, we compare instead with established results for the Heisenberg spin glass on the simple cubic lattice. Since the number of nearest neighbours is the same on both lattices, we expect that this comparison will be adequate to establish the trend in $T_{\rm f}/\Delta$ with $J$. For the cubic lattice with Gaussian nearest neighbour exchange of zero mean and unit variance, $T_{\rm f} = 0.129$.\cite{conventional}  Assuming that it is appropriate to compare our rectangular exchange distribution with a Gaussian by simply equating variances, we conclude that large $J$ increases the value of $T_{\rm f}/\Delta$ by the factor $\sqrt{3} \times 0.23  /0.129 \approx 3.1$. This substantial effect is physically reasonable: thermal fluctuations in a geometrically frustrated system are much more restricted than in a conventional spin glass. The spin glass phase therefore extends to higher temperatures than in a system with average exchange $J=0$. Turning to exponent values, our result for $\nu$ lies within the range ($1.01-1.50$) reported from simulations of the conventional Heisenberg spin glass on the simple cubic lattice. \cite{conventional} Comparison with experimental results for the pyrochlore antiferromagnet $\ymoo$ is also possible. The experimental values are $\gamma=2.9(5)$ and $\beta=0.8(2)$ \cite{Gingras1997}.  Our value of $\gamma$ (given above) is significantly smaller than the experimental one, while the result we obtain for $\beta$ using hyperscaling,  $\beta=0.8(3)$, is in close agreement. Experience from very large scale simulations of conventional three-dimensional Heisenberg spin glasses unfortunately  suggests that a much bigger computational effort than our own would be necessary to characterise critical behaviour reliably. We nevertheless have confidence in our central result from simulation: that weak exchange randomness in a geometrically frustrated magnet induces a spin glass transition, with a higher transition temperature than would be the case of the mean exchange were not strongly antiferromagnetic. 

\section{Conclusions}
\label{conclusions}


Spin freezing has been observed experimentally at low temperatures in many geometrically frustrated magnetic materials, and its origin has long been unclear. In this work we have carried out a detailed analytical and numerical study of models with weak disorder, showing that this is a mechanism that produces freezing and a low temperature spin glass phase.

Analytically, we have studied geometrically frustrated antiferromagnets perturbed by weak exchange randomness. This form of disorder may be generated by random strains in the sample, arising from non-magnetic chemical disorder.  We have shown that it leads to a spin glass phase at low temperatures.   The main result, expressed by Eqns. \eqref{th-ice-th} and \eqref{th-vec-th}, is that a model of the Edwards-Anderson (EA) type, but with dominant, mean geometrically frustrated antiferromagnetic exchange, falls into the same universality class as the usual EA model.\cite{Edwards-Anderson1975} We find no essential deviations from conventional spin glass results for the character of the transition to the low temperature phase. However, the (mean-field) transition temperature is increased by the long-range correlations of the pure antiferromagnet, by a factor of $\sqrt{2}$ as compared to a system with no mean antiferromagnetic interaction.

In a complementary approach, we have shown qualitatively how a similar conclusion arises in a model with dilute disorder. For this case we map the initial system, consisting of spins on a regular, frustrated lattice with a low concentration of disordered interactions, to an effective system, made up of pseudospins at random sites with entropic interactions that are dipolar in character. Such interactions, although long range, are expected to yield a conventional spin glass transition,\cite{Bray-Moore1982} as we find for homogeneous disorder.

In Sec \ref{ns} we have checked these ideas using numerical simulations. For conventional Ising Edwards-Anderson systems it has long been accepted that there is a finite temperature spin glass transition in three dimensions. There is also compelling evidence for spin glass ordering in real experimental three-dimensional spin glasses. The existence of a spin glass transition in three-dimensional Heisenberg Edwards-Anderson spin glasses has been controversial, but the most recent exhaustive studies \cite{conventional} indicate that such a transition does occur.  The data we present  supports the corresponding conclusion that spin freezing transitions also occur in disordered geometrically frustrated antiferromagnets. A disputed issue for three-dimensional Heisenberg Edwards-Anderson spin glasses is whether the spin-glass transition is distinct from a chiral ordering transition.\cite{conventional} We have not attempted to address this for geometrically frustrated systems
but we would not expect any qualitative differences from conventional spin glasses without geometrical frustration.

While our simulations have been of a three-dimensional system and the replica treatment we have presented has been at a mean-field level, some geometrically frustrated antiferromagnets exhibiting spin freezing, including {$\rm{SrCr_{8}Ga_{{4}}O_{19}}$}, are in fact quasi-two dimensional. Since our central conclusion is that universal features of spin glass ordering in geometrically frustrated magnets with weak quenched disorder should be the same as those in conventional
 Edwards-Anderson systems, and since the two-dimensional Heisenberg Edwards-Anderson model is believed not to have spin-glass ordering,\cite{binder1986} it is necessary to appeal to weak interlayer coupling or spin anisotropy to account for spin freezing in quasi-two dimensional geometrically frustrated Heisenberg systems; note that the relevant scale for the interlayer
coupling to affect spin glass ordering is set by $\Delta$ rather than J.


Our numerical simulations support the theoretical expectation that the critical temperature is proportional to $\Delta$ for $J \gg \Delta$, with $T_f \simeq 0.23 \Delta \simeq 0.40 \tilde{\Delta}$. The observed values of $T_f$ therefore imply fluctuations in exchange interactions strength with a variance that is a few percent of the mean in  $\rm{SrCr_{8}Ga_{{4}}O_{19}}$ and some ten times larger in $\ymoo$. A direct experimental search for such exchange fluctuations would be of great interest.


\begin{acknowledgments}
We thank C. Castelnovo, P. C. W. Holdsworth and R. Moessner for helpful discussions. This work was supported by EPSRC Grant No. EP/D050952/1.
\end{acknowledgments}


\begin{thebibliography}{99}
\bibitem{footnote2} This factor is equal to $\sqrt{2}$ with good precision, but is not
        exact, even at the mean field level, because of the approximation
        used to evaluate $\widehat{G_{\kappa\eta}^2}(0)$.
\end{thebibliography}
\end{document}